\documentclass[aps,12pt,superscriptaddress,amsfonts,amssymb,amsmath]{revtex4}
\pdfoutput=1

\usepackage{graphicx}
\usepackage{epsfig}
\usepackage{makeidx}
\usepackage{epstopdf}
\usepackage{natbib}
\usepackage{xcolor}
\usepackage{subcaption}
\usepackage{amsmath}

\begin{document}

\title{Quantum metrics with very low action in $R+R^2$ gravity}

\author{G.\ Modanese \footnote{Email address: giovanni.modanese@unibz.it}}
\affiliation{Free University of Bozen-Bolzano \\ Faculty of Science and Technology \\ I-39100 Bolzano, Italy}

\linespread{0.9}

\begin{abstract}

We have run numerical simulations of Euclidean lattice quantum gravity for metrics which are time-independent and spherically symmetric. The radial variable is discretized as $r=hL_{Planck}$, with $h=0,1,...,N$ and $N$ up to $10^5$. The Lagrangian is of the form $\sqrt{g}(R+\alpha R^2)$ (in units $c=\hbar=G=1$) and the action is positive-definite, allowing the use of a standard Metropolis algorithm with update probability $\exp(-\beta \Delta S)$. By minimizing the $R+R^2$ action with respect to conformal modes, Bonanno and Reuter have recently found analytical evidence of a non-trivial ``rippled'' ground state resembling a kinetic condensate of QCD. Our simulations at low but finite temperature ($T=\beta^{-1}$) also display strong localized oscillations of the metric, whose total action $S$ remains $\ll \hbar$ thanks to the indefinite sign of $R$. The average metric $\langle g_{rr} \rangle$ is significantly different from flat space. The scaling properties of $S$ and $\langle g_{rr} \rangle$ are investigated in dependence on $N$ and $\beta$.

\end{abstract}

\maketitle

\section{Introduction}

Since the Euclidean Einstein action of the gravitational field is not limited from below, the minimal Lagrangian $\sqrt{g}R/16\pi G$ has often been ``stabilized`` with a term $\sqrt{g}R^2$ which does not have sizable effects at macroscopic distances, but only on a very small scale \cite{stelle1977renormalization,rovelli2004quantum,hamber2008quantum,capozziello2011extended}. The $R^2$ term can spoil the unitarity of perturbation theory, but non-perturbative approaches have the potential to solve this problem \cite{hamber2009quantum,ambjorn2012nonperturbative,hamber2019vacuum,ambjorn2019towards}. Moreover, in the Asymptotic Safety scenario (with renormalization around an UV fixed point) terms quadratic in the curvature arise in a natural way \cite{niedermaier2006asymptotic,reuter2018quantum}.

Bonanno and Reuter have recently studied through non-perturbative analytical methods the continuum quadratic theory, limited to conformal modes, and found indications of a ``rippled`` ground state that violates translational symmetry \cite{bonanno2013modulated,bonanno2019structure}. They have suggested that the vacuum of asymptotically safe gravity theories quadratic in the curvature has the form of a kinetic condensate similar (if confirmed in the full theory with all physical degrees of freedom) to the Savvidy vacuum of Quantum Chromodynamics \cite{lauscher2000rotation,branchina1999antiferromagnetic}.

In this work we consider a dimensional reduction of the metric, employed in several classical and quantum gravitational models, where the only variable is the metric component $g_{rr}(r)\equiv A(r)$. The metric is supposed to be independent from time and from the angle variables; we also suppose that $g_{00}(r)=const$.

Technically it would be possible to introduce a dependence on time and on the angles and a dynamical $g_{00}$, but the chosen reduction has the advantage of leading to a strong simplification of the action. The Einstein term, here denoted as $S^R$, while $S^{R^2}$ denotes the $R^2$ term, is written simply as \cite{weinberg1973gravitation,modanese2007vacuum,modanese2019metrics,modanese2020quantum}
\begin{equation}
    S^R=\frac{\tau}{16\pi G} \int_0^\infty dr \, \sqrt{|A|} \left( \frac{rA'}{A^2}+1-\frac{1}{A} \right)
    \label{SR}
\end{equation}
The time interval $\tau$ can be interpreted as the duration of the vacuum fluctuations of the metric that we are going to simulate. These fluctuations are considered with respect to the classical vacuum solution $A(r)=1$, for which $S^R=0$. We have chosen in (\ref{SR}) natural units such that $\hbar=c=1$, and in the following we shall also fix the length unit to $L_{Pl}$ and thus $G=1$. In the numerical simulations we discretize the variable $r$ on a finite interval $(0,L)$, with the boundary condition $A(L)=1$, and we investigate the scaling in $L$.

In Sect.\ \ref{discretized-action} we write the discretized version of the action $S^R+S^{R^2}$ with variables $A_h=A(h\cdot L_{Pl})$, where $h=0,1,...,N$, so that $L=NL_{Pl}$. 

In Sect.\ \ref{results} we report the results of simulations in which this systems starts from a flat-space configuration with $A_h=1$ for all $h$, and then at each step one of the $A_h$ is randomly chosen and varied as $A_h \to A_h \pm \varepsilon$. The variation is accepted or rejected with a standard Metropolis criterion \cite{newman1999monte}, namely always accepted if $\Delta S<0$, or otherwise accepted if $\exp(-\beta \Delta S)<\xi$, being $\xi$ a random variable in the interval $(0,1)$.

In equilibrium metric configurations the variables $A_h$ oscillate quite strongly, with amplitudes which of course decrease at large $\beta$ ($\beta$ has the role of inverse temperature in the equivalent statistical system). These oscillations, however, are such that the total action remains $\ll \hbar$, thanks to the fact that the $R$ term in the action has indefinite sign and so opposite fluctuations of the $A_h$'s can compensate each other. This feature is unique of the gravitational action, even when stabilized with the $R^2$ term, and apparently leads to the formation of vacuum fluctuations much larger than in other quantum field theories \cite{modanese2000paradox}. Numerical simulations of this effect have been already reported in \cite{modanese2019metrics,modanese2020quantum}, but without the stabilizing $R^2$ term, and were therefore less clear and reliable.

The average total action scales quite exactly in proportion to $\beta^{-1}$ and $N$. This allows to extend the validity of simulations to large $N$, i.e., to a scale much larger than the Planck scale, and to a continuum limit, provided $\beta$ is increased, implying a lower temperature and thus a better approximation of the ground state. In taking the large-$N$ limit one must also consider the time duration $\tau$ of the fluctuations, which needs to be large enough to represent in a consistent way the adiabatic switch on/off of the fluctuations. Otherwise, the action should be modified by including the contributions to $R$ coming from time derivatives.

Among the quantities measured in the simulations, the most interesting ones are probably the averages $\langle A_h \rangle=\langle g_{rr}(hL_{Pl}) \rangle$. These show what average metric emerges spontaneously from the action without imposing a background. In long runs $\langle A_h \rangle$ turns out to be independent from $h$ (metric constant in the interval $(0,L)$) and significantly different from 1. Let us define $\psi_h=\langle A_h -1\rangle $, and let $\psi$ be the lattice average of $\psi_h$, namely
\begin{equation}
    \psi=\frac{1}{N+1} \sum_{h=0}^N \psi_h
    \label{def-psi}
\end{equation}
while $\sigma_\psi$ quantifies the (small) spatial fluctuations of $\psi$:
\begin{equation}
    \sigma^2_\psi=\frac{1}{N+1} \sum_{h=0}^N (\psi_h-\psi)^2
\end{equation}
The ``order parameter`` $\psi$ turns out to be independent from $\alpha$ and $\beta$ and to scale in $N$ as $N^{-1}$ (Tab.\ \ref{table2}).

The measured difference between the average metric and flat space may look tiny (after all, we are close to the Planck scale), but it is relevant, in our opinion, if correctly interpreted. In fact the scale-independent product $M=N\psi$ turns out to be equal to the Planck mass and can be regarded as twice the total virtual mass of the vacuum fluctuation described by the average metric $\psi_h$, as discussed in Sect.\ \ref{interp}.

Sect.\ \ref{concl} contains our conclusions.

\section{The discretized $R+R^2$ action}
\label{discretized-action}

Let us choose units in which not only $\hbar=c=1$ (natural units, as in \cite{modanese2020quantum}), but also $G=1$. This means that all lengths are measured as multiples of the fundamental length $L_{Pl}$. We consider metric configurations that differ from flat space in an interval of $r$ from 0 to $L$ and divide this interval into $N$ parts in the discretized calculation. In order to obtain a good precision, $N$ should be as large as possible; the data and pictures of Sect.\ \ref{results} are mostly for $N=1600$, but the scaling in $N$ is also very important, and we have data up to $N\sim 10^5$. The discretization cut-off is chosen as $d=L/N= 1=L_{Pl}$. This means that the interval $L$ has a length that is $N$ times the Planck length. For the time duration $\tau$ of our fluctuations we fix at first $\tau=1$; this is however a very short time and if we want to increase $N$ (and therefore $L$) in a way that is consistent with our quasi-stationary approximation for the metric, we will eventually need to increase $\tau$. We shall see that this is possible thanks to the favourable scaling of the simulation in the inverse temperature $\beta$.

The discretized action for time-independent metrics with spherical symmetry and $g_{00}=const.$ can be written as
\begin{equation}
    S=\sum_{h=0}^N S_h = \tau d \sum_{h=0}^N \left( S_h^R+S_h^{R^2} \right)
    \label{total-action}
\end{equation}
where the term $S_h^R$ is given by
\begin{equation}
    S_h^R=\sqrt{|\hat{A}_h|} \tilde{S}_h
\end{equation}
\begin{equation}
    \tilde{S}_h=\frac{1}{\hat{A}_h^2} \left( \frac{A_{h+1}-A_h}{d}\right) \left( h+\frac{1}{2} \right) d +1-\frac{1}{\hat{A}_h}
\end{equation}
and 
\begin{equation}
    \hat{A}_h=\frac{1}{2}(A_{h+1}+A_h)
\end{equation}

It is straightforward to see that $S^R$ is the discretized version of the continuum action (\ref{SR}), since $\hat{A}_h$ is the average of $A$ on the $h$-th small space interval, $(A_{h+1}-A_h)/d$ is an estimate of $A'$ on the same interval and $r$ is discretized as $(h+1/2)d$. 

The contribution of the continuum term $\alpha \sqrt{g}R^2$ is obtained by taking the square of $\tilde{S}_h$ and dividing by $r^2$:
\begin{equation}
    S_h^{R^2}=\frac{\alpha \sqrt{|\hat{A}_h|}}{d ^2 \left( h+\frac{1}{2} \right)^2} \tilde{S}_h^2
\end{equation}

The resulting total action (\ref{total-action}) is relatively simple, considering that we are dealing with a gravitational theory quadratic in the curvature.
In the term $S^{R^2}$, $\alpha$ is an adimensional coupling whose amplitude will be set empirically in the simulations following certain criteria explained below. Our definition of $\alpha$ differs from that in \cite{bonanno2013modulated,bonanno2019structure} by a factor $4\pi$.

\subsection{Description of the Metropolis algorithm}
\label{metropolis}

The code of the Monte Carlo algorithm, both in Python and C, is appended at the end of the source of the arXiv preprint of this work.
In the code, $m$ is the number of sub-cycles, usually 200; at the end of each sub-cycle, the algorithm visualizes the current value of the total action $S$, the latest variations of $S$ (in the parts $R$ and $R^2$), and the current value of the ratio between the steps with $\Delta S$ positive and negative, which indicates the thermalization status.
$n$ is the number of elementary steps in a sub-cycle, typically from 20 to 80 millions. At each elementary step one of the $N$ variables $A_h$ is randomly chosen and varied as $A_h\pm \varepsilon$, with $\varepsilon=10^{-6}$.

The variable \texttt{sommaA} contains the sum of the values of $A_h$ used at the end to compute the average $\langle A_h \rangle$. It is updated only when $A_h$ changes, and keeps track of the number of steps without changes through the counter \texttt{contA}. On the other hand, the corresponding sums for $\exp(-\beta \Delta S)$, $S$ and $S^2$ and the counters for $\Delta S$ positive (accepted and not) and $\Delta S$ negative are updated at each step. All the averages are computed excluding an initial equilibration time corresponding to $m_0<m$ sub-cycles (normally $m_0=100$).

When $A_h$ is changed, the action changes in two of its terms $S_h$; the first one involves $A_h$ and $A_{h+1}$, and for it the quantity $\tilde{S}_h$ is written as

\texttt{Shtilde=(Ath2*(h+0.5)*(A[h+1]-A[h])+1-Ath1)}

\noindent
where we evaluate in advance 

\texttt{Ath=0.5*(A[h+1]+A[h])}, \ 
\texttt{Ath1=1/Ath}, \ \texttt{Ath2=Ath1*Ath1}

The second term involves $A_h$ and $A_{h-1}$ and for it the quantity $\tilde{S}_h$ is written as

\texttt{ShtildeM=(AthM2*(h-0.5)*(A[h]-A[h-1])+1-AthM1)}

\noindent
with

\texttt{AthM=0.5*(A[h]+A[h-1])}

\noindent
etc.

The two changing terms of the action are computed (including suitable factors for $S_h^R$ and $S_h^{R^2}$) with the old value of $A_h$, then with the new one and finally one takes the difference $\Delta S$.

\section{Results}
\label{results}

\subsection{Runs with fixed $N$ (scaling in $\beta$)}
\label{fixed-N}

\begin{figure}[h]
  \begin{center}
\includegraphics[width=7.0cm,height=5.1cm]{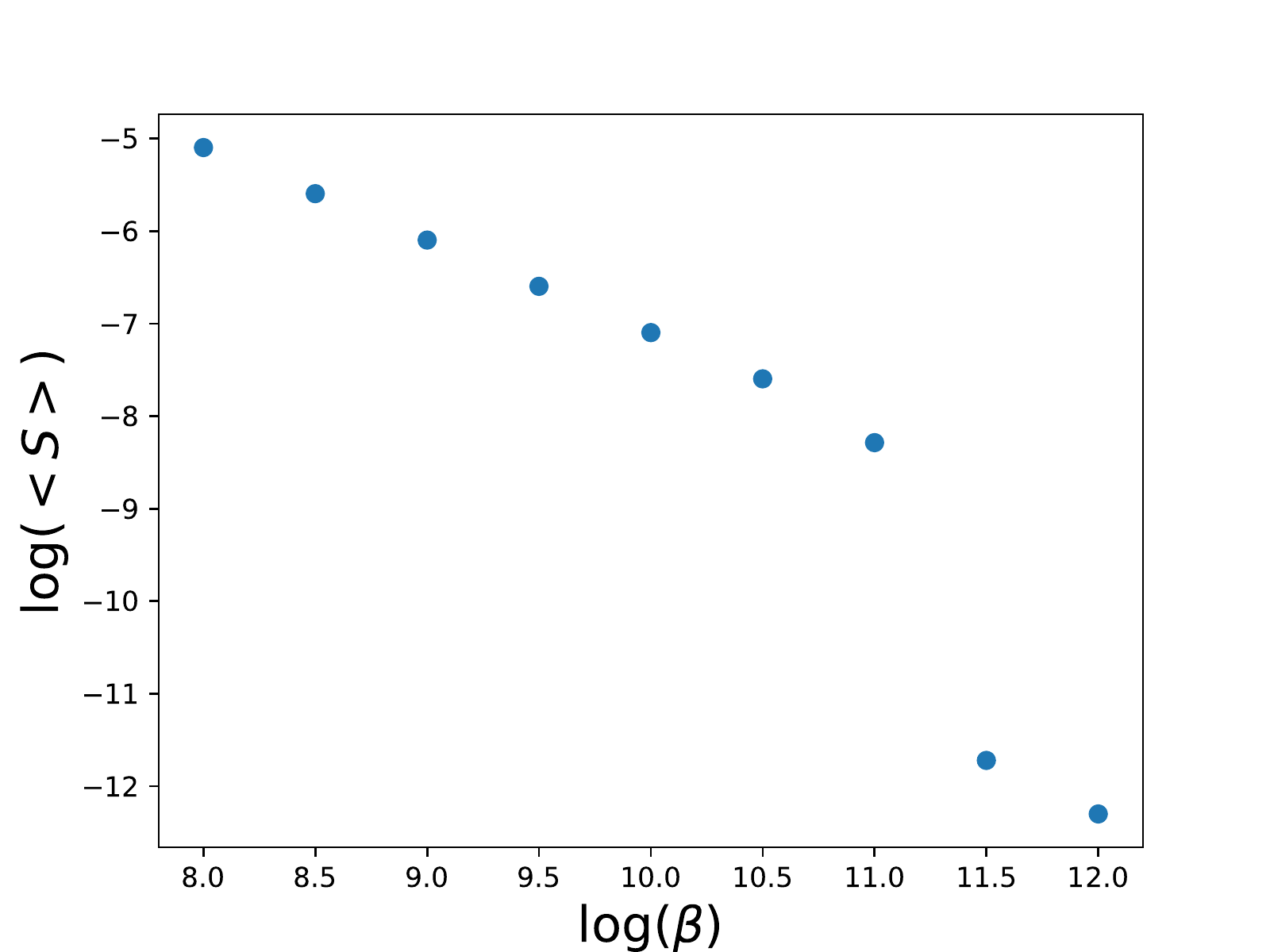}
\caption{ Log-log plot of the average action in dependence on $\beta$, with $\alpha=25$, $N=1600$ (see complete data in Tab.\ \ref{table1}). $\langle S \rangle$ decreases almost exactly as $\beta^{-1}$ up to $\beta \simeq 10^{11}$, then it drops abruptly, mixing with the noise.
} 
\label{grafico-azione-beta}
  \end{center}
\end{figure}

In these runs the number of lattice spacings is kept constant at $N=1600$. The average values obtained for the action at equilibrium by changing the inverse temperature $\beta$ are summarized in Tab.\ \ref{table1} and in the plot of Fig.\ \ref{grafico-azione-beta}, where $\langle S \rangle$ is first seen to decrease as $\beta^{-1}$ and then to collapse below the noise level for large $\beta$.

\begin{figure}[h]
\begin{subfigure}{.5\textwidth}
    \includegraphics[width=7.0cm,height=5.1cm]{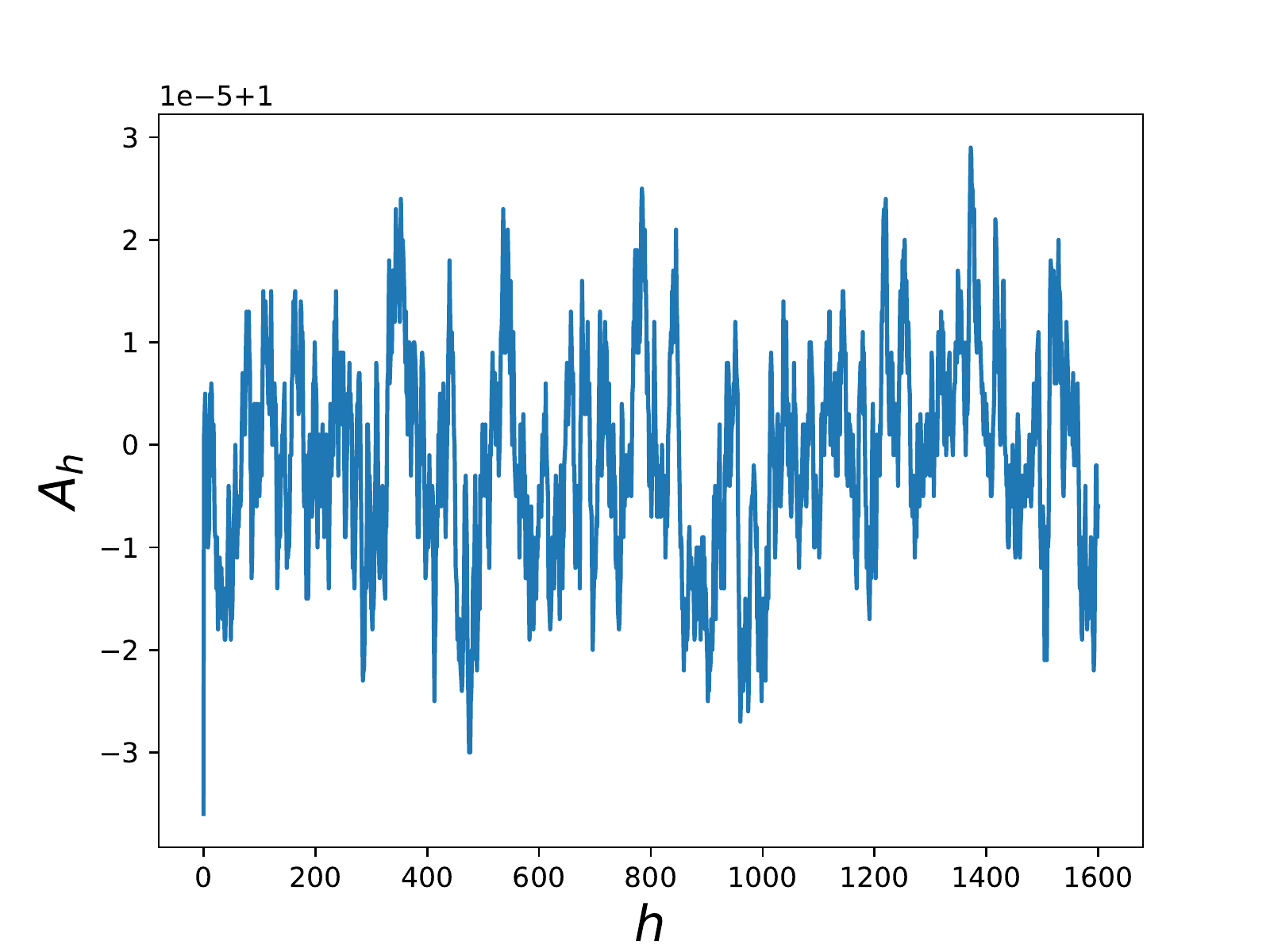}
  \caption{}
\end{subfigure}
\begin{subfigure}{.5\textwidth}
    \includegraphics[width=7.0cm,height=5.1cm]{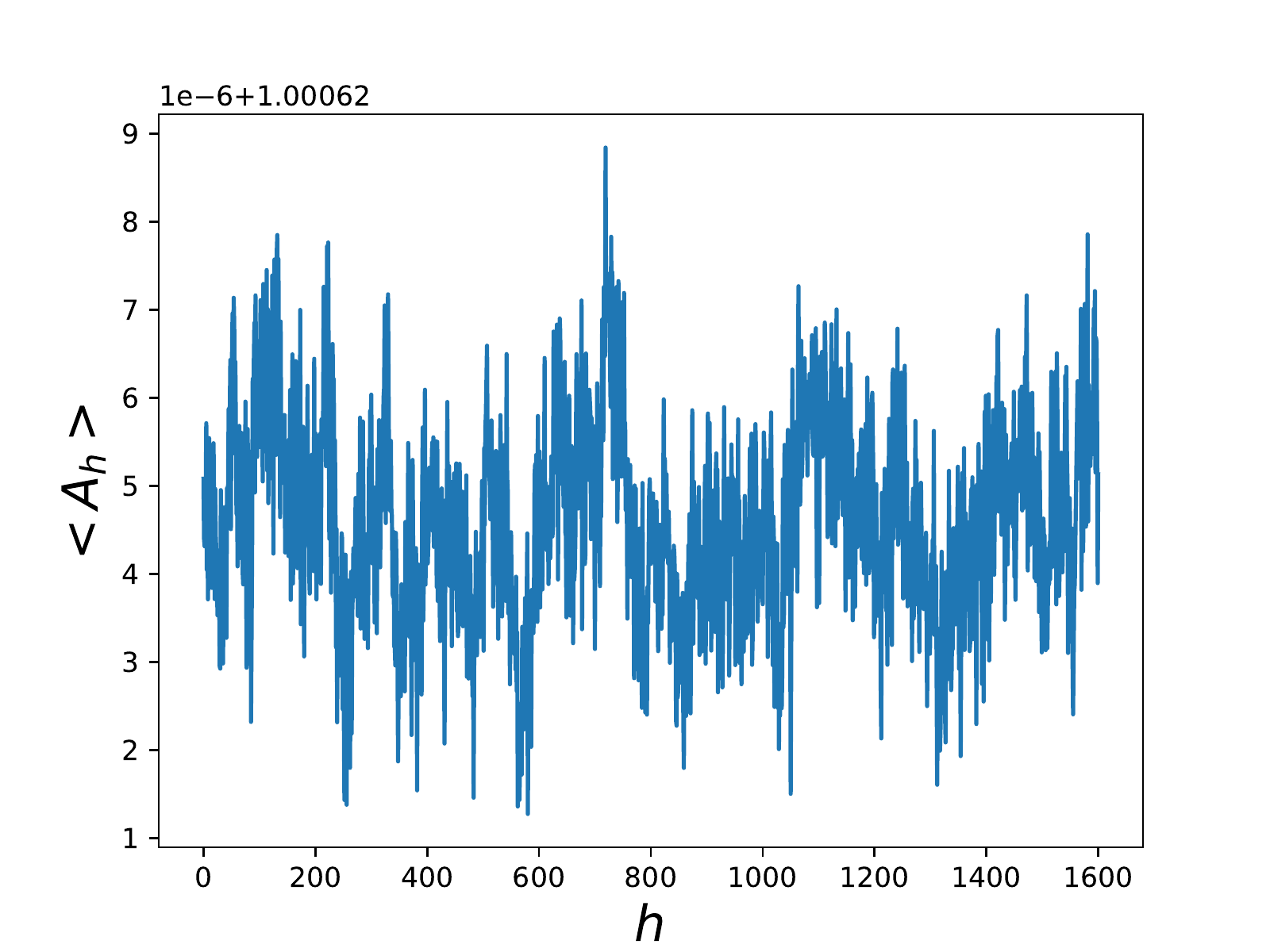}
  \caption{}
\end{subfigure}
\begin{subfigure}{.5\textwidth}
    \includegraphics[width=7.0cm,height=5.1cm]{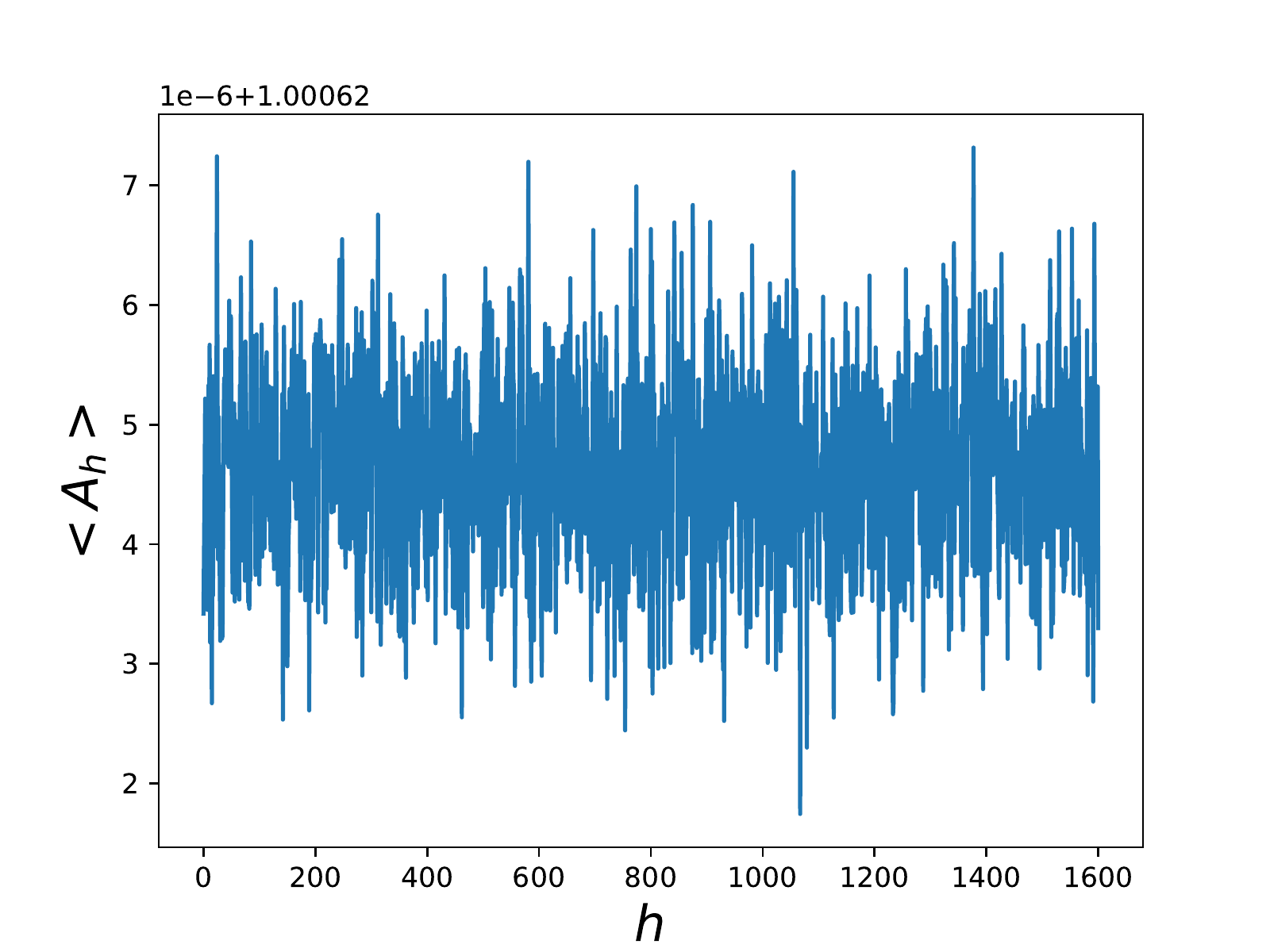}
  \caption{}
\end{subfigure}
\caption{{\bf (a)} An example of instant values of $A_h$ at equilibrium with $\beta=10^9$, $N=1600$, $\alpha=25$. {\bf (b)} Average values $\langle A_h \rangle$ with $\beta=10^9$, $N=1600$, $\alpha=25$, 200 cycles of $10^7$ steps. $\langle A_h \rangle$ is almost independent from $h$ and equal to 1.000625, with fluctuations of the order of $10^{-6}$. {\bf (c)} Average values $\langle A_h \rangle$ with $\beta=3.16\cdot 10^{11}$, $N=1600$, $\alpha=25$, 200 cycles of $10^7$ steps. Fluctuations are smaller than in (b). The metric is practically constant and different from flat space.
} 
\label{Ah-foto-media}
\end{figure}

Fig.\ \ref{Ah-foto-media} illustrates the behavior of $A_h$, with an instant picture of a typical fluctuating configuration (\ref{Ah-foto-media}-(a)) and the behavior of $\langle A_h \rangle$ at two different temperatures (\ref{Ah-foto-media}-(b) and \ref{Ah-foto-media}-(c)). When the temperature is sufficiently low the average metric $\langle A_h \rangle$ becomes independent from $h$, i.e.\ from the radial coordinate, and stabilizes with small fluctuations at 1.000625, thus with an ``order parameter`` $\psi=6.25\cdot 10^{-4}$ with respect to flat space. In Sect.\ \ref{scaling-N} we show how this number scales in $N$, in such a way to keep the product $N\psi$ equal, with a good precision, to the Planck mass.

One of the challenges posed by the numerical simulations is the choice of ``good`` values of the coupling $\alpha$ and inverse temperature $\beta$. Since for weak fields one normally has $R^2 \ll R$, it seems at first natural to choose a large value of $\alpha$ in order to have an effective stabilization. However, by analysing in several trials the random variations of the terms $R$ and $R^2$, it is seen that the term $R$ becomes almost irrelevant if $\alpha$ is much larger than $\sim 10^3$, and in this case the thermalization of the algorithm is difficult to achieve. On the other hand, if $\alpha$ is less than $\sim 10^1$ the stabilization is not sufficient and the system may collapse as a consequence of some large negative fluctuations of the action, especially if $\beta$ is not large enough. For this reason we set $\alpha=25$ in most of the simulations. Changes in $\alpha$ do not affect the values of $\langle S \rangle$ or $\langle A_h \rangle$, but only the transition probability in the algorithm (see an example in Tab.\ \ref{table1}).

When the algorithm has reached equilibrium, the number of accepted steps with $\Delta S>0$  is equal to a fraction $\langle e^{-\beta \Delta S} \rangle$ of all steps with $\Delta S>0$. For example, when $\beta=10^8$ we can deduce from Tab.\ \ref{table1} that out of 192 random variations of $A_h$ there will be on average 92 with $\Delta S<0$ and 100 with $\Delta S>0$, of which 92 accepted. When the temperature drops to $\beta=10^{11}$, out of 106 variations there are 6 with $\Delta S<0$ and 6 accepted with $\Delta S>0$, etc.

\begin{table}
\begin{center}
\begin{tabular}{|c|c|c|c|} 
\toprule
$\beta$ & $\langle S \rangle$ & $\langle e^{-\beta \Delta S} \rangle$ & $\sigma_S/S$ \nonumber \\
\hline
$10^8$ & $8.00\cdot 10^{-6}$ & 0.92 & 0.04 \nonumber \\
$3.16\cdot 10^8$ & $2.54\cdot 10^{-6}$ & 0.86* & 0.04 \nonumber \\
$10^9$ & $8.00\cdot 10^{-7}$ & 0.78 & 0.04 \nonumber \\
$3.16\cdot 10^9$ & $2.53\cdot 10^{-7}$ & 0.64 & 0.04 \nonumber \\
$10^{10}$ & $8.00\cdot 10^{-8}$ & 0.45 & 0.04 \nonumber \\
$3.16\cdot 10^{10}$ & $2.53\cdot 10^{-8}$ & 0.25 & 0.04 \nonumber \\
$10^{11}$ & $5.17\cdot 10^{-9}$ & 0.06 & 0.07 \nonumber \\
$3.16\cdot 10^{11}$ & $1.9\cdot 10^{-12}$ & $2.7\cdot 10^{-4}$ & 2 \nonumber \\
$10^{12}$ & $5.01\cdot 10^{-13}$ & $2.2\cdot 10^{-4}$ & 2 \nonumber \\
\hline
\end{tabular}	
\caption{Dependence on $\beta$ (inverse temperature) of the average action $\langle S \rangle$, the transition probability $\langle e^{-\beta \Delta S} \rangle$ and the ratio $\sigma_S/\langle S \rangle$. The lattice size $N$ is fixed at 1600, $\alpha=25$, the number of simulation steps varies between $4\cdot 10^9$ and $8\cdot 10^9$. If we change $\alpha$ (coupling to $R^2$), only the transition probability changes: for example, when we have $\langle e^{-\beta \Delta S} \rangle=0.86$ with $\alpha=25$ (*), we would instead obtain 0.75 with $\alpha=100$ and 0.67 with $\alpha=200$; i.e., increasing $\alpha$ has a similar effect to a diminution in the temperature, but only upon the transition probability, not on the average of $S$ and also not on the average of $A_h$, equal for all $\alpha$ and $\beta$ to 1.000625, with fluctuations $\sim 10^{-5}$ for $\beta=10^9$ and decreasing in proportion to $\sqrt{\beta}$ (see Fig.\ \ref{Ah-foto-media}).
}		
\label{table1}
\end{center}
\end{table}

\subsection{Scaling in $N$}
\label{scaling-N}

If we keep the temperature fixed and increase $N$, we can observe the scaling in $N$ of the average action and of the ``order parameter`` $\psi$ defined in (\ref{def-psi}) (the lattice average of $\langle A_h-1 \rangle$). Some results are summarized in Tab.\ \ref{table2}. The average action scales exactly as $N$, while $\psi$ scales as $N^{-1}$. The product $\psi N$ is equal to 1, within errors, and is independent from $\alpha$ and $\beta$. This product can be interpreted as a mass, as discussed in Sect.\ \ref{interp}, and thus it is equal to 1 Planck mass (about $2\cdot 10^{-8}$ Kg). This is a remarkable coincidence which probably has some fundamental explanation, but emerges here from long numerical simulations which proceed directly from the choice of the action (\ref{total-action}) without any other assumption. We should stress, however, that the value 1 for the product might change in case of different conventions on the units, becoming equal for instance to $2\pi$ or $(2\pi)^{-1}$.

\begin{table}
\begin{center}
\begin{tabular}{|c|c|c|c|} 
\toprule
$N$ & $\psi$ & $\sigma_\psi/ \psi$ & $\psi \cdot N$ \nonumber \\
\hline
1600 & $6.25\cdot 10^{-4}$ & $1\cdot 10^{-3}$ & 1.000 \nonumber \\
3200 & $3.12\cdot 10^{-4}$ & $2\cdot 10^{-3}$ & 0.999 \nonumber \\
6400 & $1.56\cdot 10^{-4}$ & $7\cdot 10^{-3}$ & 1.00 \nonumber \\
12800 & $7.8\cdot 10^{-5}$ & $2\cdot 10^{-2}$ & 1.00 \nonumber \\
25600 & $3.9\cdot 10^{-5}$ & $5\cdot 10^{-2}$ & 1.00 \nonumber \\
\hline
\end{tabular}	
\caption{Values of the order parameter $\psi$ (lattice average of $\langle A_h-1 \rangle $) in dependence on the size $N$ of the lattice.
}		
\label{table2}
\end{center}
\end{table}

\subsection{Role of the boundary condition}
\label{role}

All the results above were obtained with the fixed boundary condition $A_{N+1}=1$. (For $h=0,1,...,N$ there is an initial condition $A_h=1$, but then $A_h$ is free to fluctuate.)

We also made runs in which the boundary condition is $A_{N+1}=1+\delta$, with $-10^{-4} \leq \delta \leq 10^4$. It turns out that in this case the action remains $\ll 1$ but does not drop to such small values as in Tab.\ \ref{table1}. Note that the initial value of the discretized action with the boundary condition $A_{N+1}=1+\delta$ is not zero but $\simeq N|\delta|$, entirely due to the jump between $h=N$ and $h=N+1$. What is remarkable is that $\langle A_h \rangle$ at equilibrium does not depend on $\delta$ and is always equal to the value with $\delta=0$. In other words, the boundary condition does not affect the equilibrium order parameter, at least for the tested values of $\delta$.

\section{Possible interpretation of the product $N\psi$ as virtual mass of the fluctuations}
\label{interp}

If we write the metric component $g_{rr}(r)$ for the classical Schwarzschild solution in units in which $c=\hbar=G=L_{Pl}=1$, it takes the form
\begin{equation}
    g_{rr}^{Schw}(r)=\left( 1-\frac{r_{Schw}}{r} \right)^{-1}
    \label{grr-schw}
\end{equation}
where $r_{Schw}=2M$, and $M$ is the mass of the source. (Notice that in natural units $c=\hbar=1$ one has $r_{Schw}=2GM$, where $G$ has dimensions $l^2$ and $M$ has dimensions $l^{-1}$.) At distances $r\gg r_{Schw}$, the metric $g_{rr}^{Schw}(r)$ can be approximated as $\simeq 1+r_{Schw}/r$. This means that by measuring the distant field generated by a mass and fitting the coefficient of its $1/r$ dependence, we can in principle find the mass of the source. This mass would coincide with the ADM mass computed from a surface integral of the metric at spatial infinity.

Now, consider a field configuration which has a ``far metric`` with behavior $g_{rr}\simeq 1+r_{Schw}/r$, and suppose that when we get closer to the source we observe more precisely a metric like (\ref{grr-schw}) down to a distance $L=NL_{Pl}$. We know that the scalar curvature of the Schwarzschild metric is zero, and thus on the interval $(L,+\infty)$ the observed metric has the same action as flat space.

Next suppose to get even closer to the source. In classical gravity we expect that  ``something happens`` here, in the sense that we encounter a singularity or in any case some non-zero contribution to the action. But in the quantum theory we have found a huge set of configurations in statistical equilibrium for which the action in the interval $(0,L)$ is essentially zero and the \emph{average} metric is almost exactly constant, namely $\langle g_{rr}(r) \rangle=1+\psi$ (although the metric in each configuration has oscillations in $r$). Therefore it makes sense to impose, on average, the matching condition
\begin{equation}
    \langle g_{rr}(L) \rangle = g_{rr}^{Schw}(L)
\end{equation}
or
\begin{equation}
    1+\psi=\left( 1-\frac{2M}{NL_{Pl}} \right)^{-1}
\end{equation}
Since $N \gg 1$, for $M=1/2$ this yields $\psi \simeq N^{-1}$, in agreement with the result $N\psi \simeq 1$ found in the simulations.

In other words, a metric $g_{rr}$ which outside the interval $(0,NL_{Pl})$ behaves like $(1-1/r)^{-1}$ and inside the interval fluctuates like in the simulations, with average $1+\psi$, has action very close to zero and resembles very much, looking at its far metric, a particle of mass 1/2 in Planck units.

We thus arrive to the conclusion that the vacuum of quantum gravity contains localized fluctuations with size a multiple of $L_{Pl}$ and virtual mass equal to half the Planck mass. This idea is quite natural and not new, but in previous work singular geometries with complex topology have been considered, like wormholes or virtual black holes \cite{hawking1996virtual,preparata2000gas,garattini2002spacetime,hooft2018virtual}.
Here we have provided new numerical evidence using simpler formal ingredients. It should be added that in our simulations the size of the fluctuations can be a large multiple $N$ of the Planck length (depending also on their duration $\tau$), but the associated virtual mass is always equal to half the Planck mass, at least in the present assumption of stationary metrics with spherical symmetry and constant $g_{00}$.

\begin{figure}[h]
\begin{subfigure}{.5\textwidth}
    \includegraphics[width=7.0cm,height=5.1cm]{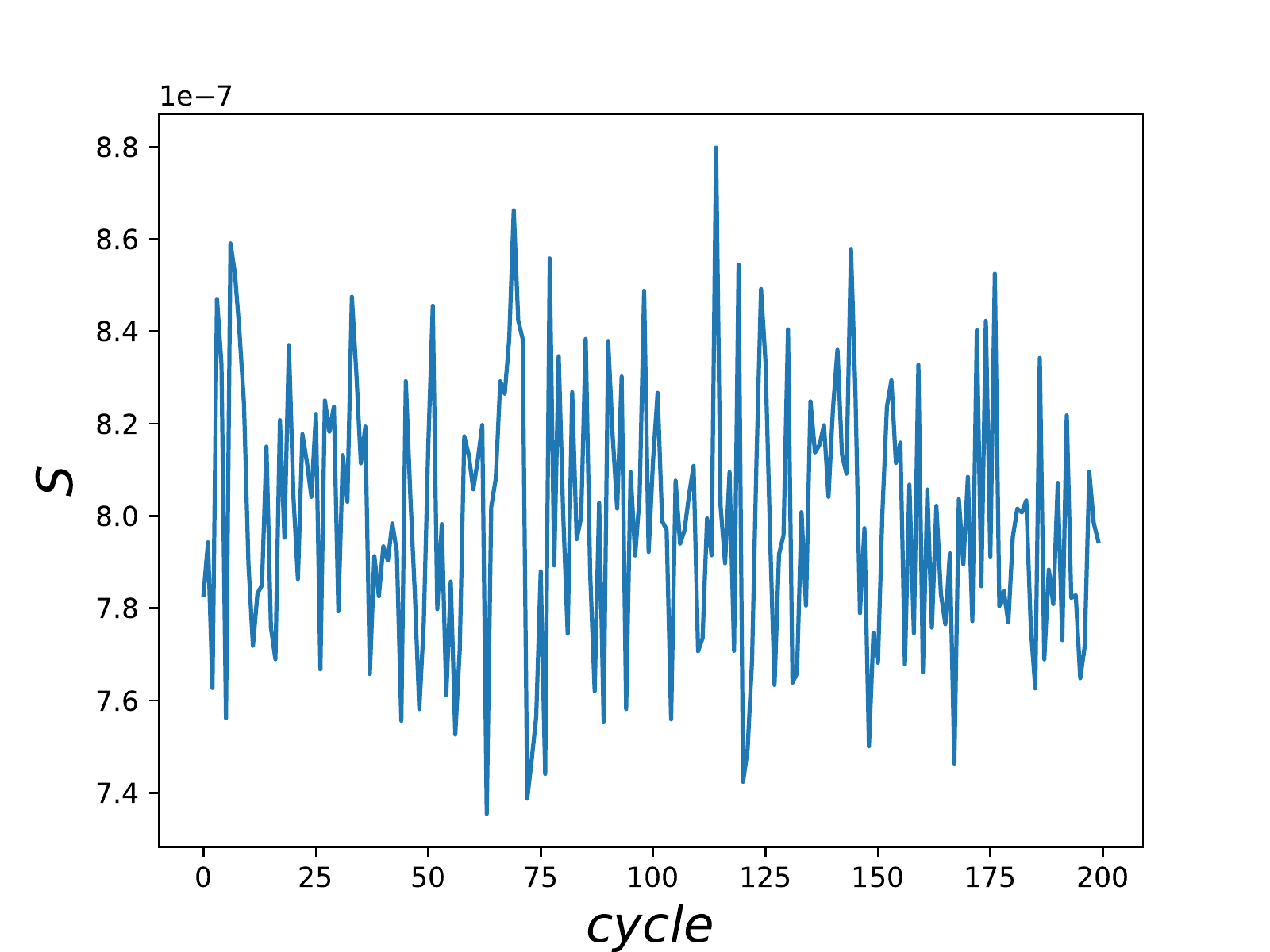}
  \caption{}
\end{subfigure}
\begin{subfigure}{.5\textwidth}
    \includegraphics[width=7.0cm,height=5.1cm]{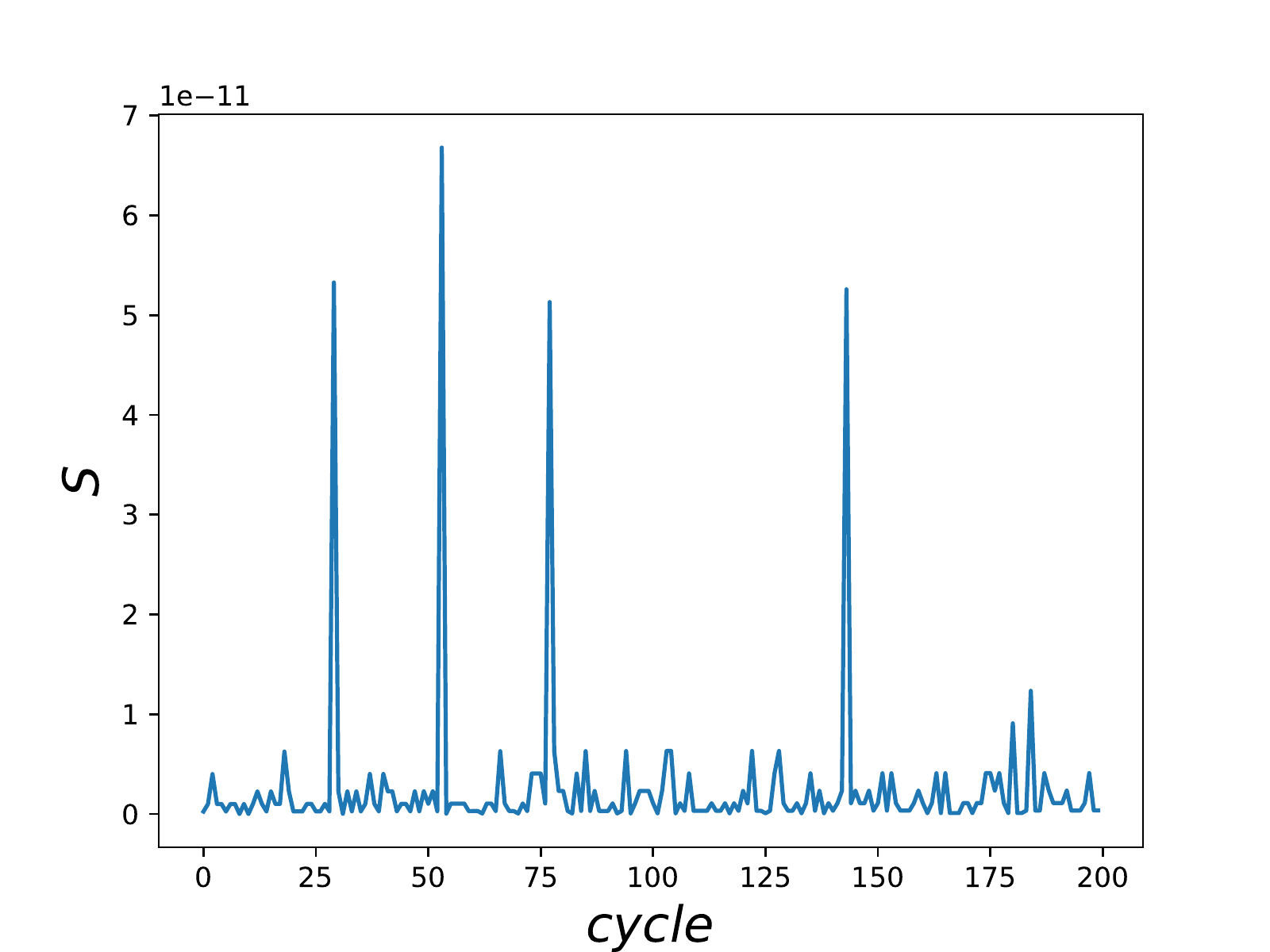}
  \caption{}
\end{subfigure}
\caption{{\bf (a)} Example of values of the action at the end of 200 sub-cycles of $2\cdot 10^7$ steps each, at middle temperature ($\beta= 10^{9}$), $N=1600$, $\alpha=25$. The noise is less than 5\% of $S$. {\bf (b)} Example of values of the action at the end of 200 sub-cycles of $2\cdot 10^7$ steps each, at low temperature ($\beta= 3.16\cdot 10^{11}$), $N=1600$, $\alpha=25$. One has $\langle S \rangle \simeq 2\cdot 10^{-12}$, $\sigma_S \simeq S$, with some fluctuations reaching $\sim 10^{-11}$.
} 
\label{azione-varie-temp}
\end{figure}

\begin{figure}[h]
    \includegraphics[width=7.0cm,height=5.1cm]{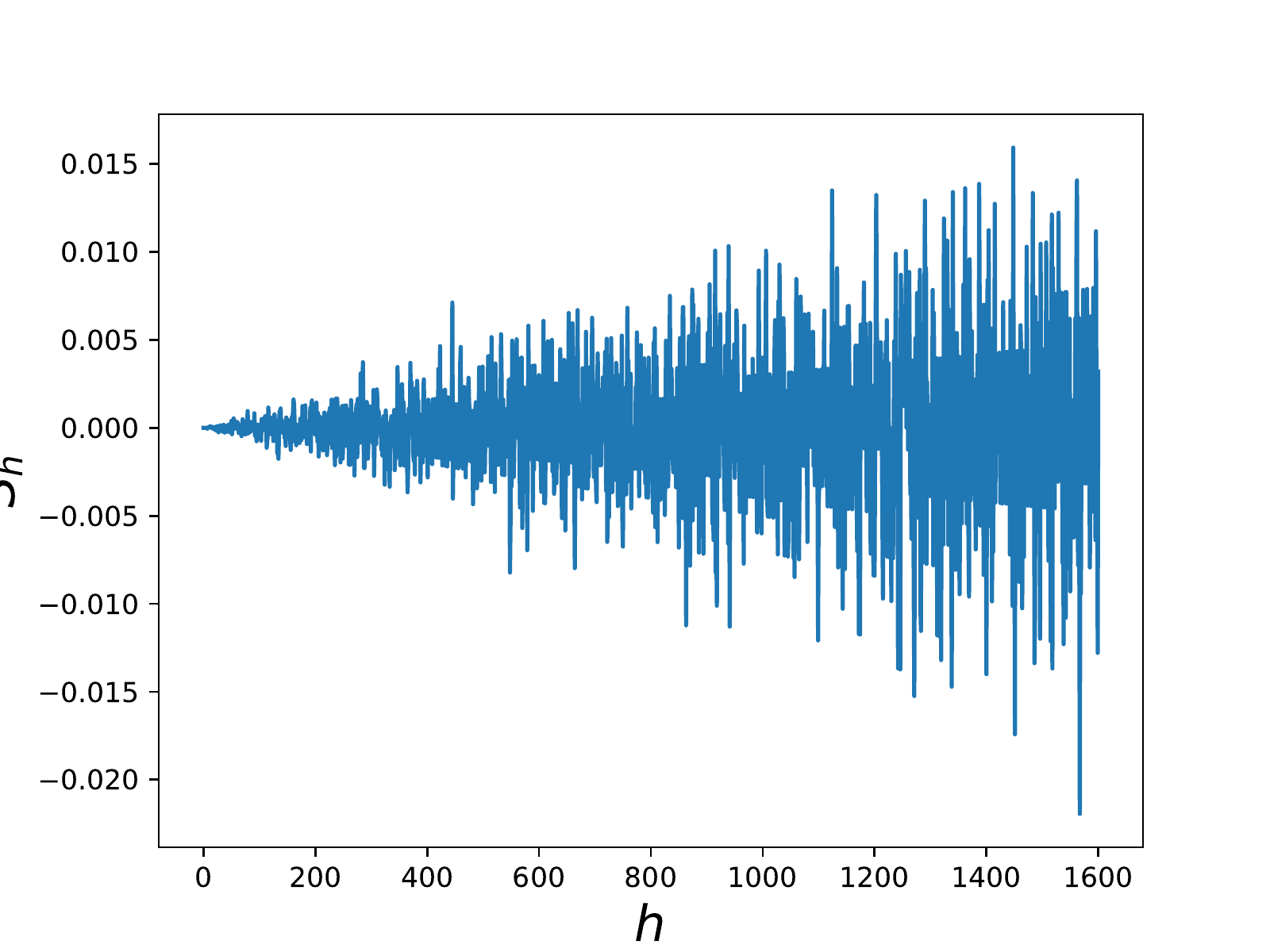}
\caption{The contributions $S_h$ to the action (\ref{total-action}) on each lattice interval, for the metric fluctuation of Fig.\ \ref{Ah-foto-media}-(a). 
The phenomenon of curvature polarization at small scale is evident: the single contributions are of the order of $\sim 10^{-2}$, but due to their randomly alternated sign they give a total action of $\sim 10^{-6}$ (Fig.\ \ref{azione-varie-temp}-(a) and Tab.\ \ref{table1}). The amplitude of the fluctuations of $S_h$ is seen to increase in proportion to $h^2$. This is probably due to the volume factor $r^2$ in the action integral.
} 
\label{foto-contributi-Sh}
\end{figure}

\section{Conclusion}
\label{concl}

In this work we have proven with reliable non-perturbative lattice simulations the existence of a class of gravitational vacuum fluctuations for which there was until now only partial numerical and analytical evidence.

In spite of the restrictions on the dynamical degrees of freedom, this result concerns the true, physical quantum gravity in 3+1 dimensions, not a toy model. The $R+R^2$ action employed is stable in the Euclidean formulation and in line with the current approaches to quantum gravity based on asymptotic safety.

One element that is missing, in comparison to more general techniques, is the explicit implementation of the diffeomorphism invariance \cite{hamber2008quantum,ambjorn2012nonperturbative}. We are essentially working in a fixed gauge, and the lattice spacing does not correspond to the physical distance; the latter could be re-obtained using the metric, either in single configurations or in an average sense.

The physical interpretation of the results is intriguing. The quantity $\psi$ that we called ``order parameter``, equal to the lattice average (on $h$) of the statistical average $\langle A_h-1 \rangle = \langle g_{rr}(hL_{Pl})-1 \rangle$, is clearly non-zero, with small fluctuations, and such that its product $N\psi$ with the size of the lattice is equal to 1 in Planck units with a precision of $10^{-3}$ to $10^{-2}$. We have discussed the physical interpretation of $N\psi$ as the virtual mass of the fluctuations, which turn out in this sense to be exactly quantized.

It is conceivable that more complex fluctuations, e.g.\ with variable $g_{00}$ or without spherical symmetry, may have a mass multiple of the minimum mass. Their numerical simulation needs a non-trivial extension of the algorithm, but appears to be feasible and will be the object of future work.

\bibliographystyle{unsrt}
\bibliography{QG2}

\end{document}